\documentclass[conference]{IEEEtran}
\IEEEoverridecommandlockouts
\usepackage{cite}
\usepackage{amsmath,amssymb,amsfonts}
\usepackage{graphicx}
\usepackage{textcomp}
\usepackage{xcolor}
\usepackage{amsmath,amssymb,amsfonts}
\usepackage{graphicx}
\usepackage{textcomp}
\usepackage{xcolor}
\usepackage{pgfplots}

\usepackage{url}
\usepackage{mathtools,algorithm,xpatch}
\usepackage{blindtext}
\usepackage{algpseudocode}
\usepackage{bm}
\usepackage{todonotes}
\usepackage{booktabs} 
\makeatletter

\def\tagform@#1{\maketag@@@{(\ignorespaces\textbf{#1}\unskip\@@italiccorr)}}
\renewcommand{\eqref}[1]{\textup{{\normalfont(\ref{#1}}\normalfont)}}
\makeatother

\definecolor{darkgreen}{rgb}{0.0, 0.5, 0.0}
\definecolor{darkred}{rgb}{1.0, 0.0, 0.22}

\usepackage{listings}
\usepackage{caption}

\DeclareCaptionFormat{listing} {\parbox{\textwidth}{\hspace{15pt}#1#2#3} } 
\definecolor{codegreen}{rgb}{0,0.6,0}
\definecolor{codegray}{rgb}{0.5,0.5,0.5}
\definecolor{codepurple}{rgb}{0.58,0,0.82}
\definecolor{backcolour}{rgb}{0.95,0.95,0.92}

\newcommand\veristyle{\lstset{
    language = Verilog,
    xleftmargin=5.0ex,
    backgroundcolor=\color{backcolour},   
    commentstyle=\color{codegreen},
    keywordstyle=\color{magenta},
    numberstyle=\tiny\color{codegray},
    stringstyle=\color{codepurple},
    basicstyle=\ttfamily\footnotesize,
    breakatwhitespace=false,
    breaklines=true,                 
    captionpos=b,
    keepspaces=true,                 
    numbers=left,                    
    numbersep=5pt,                  
    showspaces=false,                
    showstringspaces=false,
    showtabs=false,                  
    tabsize=2
}}

\renewcommand{\baselinestretch}{0.99}
\lstnewenvironment{verilog}[1][]
{
\veristyle
\lstset{#1}
}
{}

\usepackage{etoolbox}
\makeatletter
\patchcmd{\@makecaption}
  {\scshape}
  {}
  {}
  {}
\makeatother
\def\tablename{Table}

\def\BibTeX{{\rm B\kern-.05em{\sc i\kern-.025em b}\kern-.08em
    T\kern-.1667em\lower.7ex\hbox{E}\kern-.125emX}}
\pgfplotsset{compat=1.18}
\begin{document}

\title{Simopt - Simulation pass for Speculative Optimisation of FPGA-CAD flow
\thanks{979-8-3503-4959-7/24/\$31.00 ©2024 IEEE}}
\author{\IEEEauthorblockN{Eashan Wadhwa \& Shanker Shreejith}
\IEEEauthorblockA{Department of Electronic and Electrical Engineering,
Trinity College Dublin\\
Dublin, Ireland\\
\{wadhwae, shankers\}@tcd.ie}}
\maketitle

\begin{abstract}
Behavioural simulation is deployed in CAD flow to verify the functional correctness of a Register Transfer Level (RTL) design. 
Metadata extracted from behavioural simulation could be used to optimise and/or speed up subsequent steps in the hardware design flow. 
In this paper, we propose Simopt, a tool flow that extracts simulation metadata to improve the timing performance of the design by introducing latency awareness during the placement phase and subsequently improving the routing time of the post-placed netlist using vendor tools. For our experiments, we adapt the open-source Yosys flow to perform Simopt-aware placement. Our results show that using the Simopt-pass in the design implementation flow results in up to 38.2\% reduction in timing performance (latency) of the design. 
\end{abstract}


\section{Introduction}
Behavioural simulation tools play an integral role in the Electronic Design Automation (EDA) design flow, allowing the design engineer to verify the functional and logical correctness of their design under the same constraints of final implementation at the higher abstraction level. 
A complete simulation flow generates a tremendous amount of data and can provide insights into the low-level operation of the design to primarily aid in verification and validation; however, these insights are rarely (if ever) used by subsequent steps and tools in the EDA flow. 
Subsequent steps in the EDA flow typically focus on optimising the performance (or similar user-driven constraint) of the implemented design. 
In a standard Field Programmable Gate Array (FPGA) design flow, user-generated RTL designs are mapped to gate-level netlists through logic synthesis followed by iterative place and route steps to generate the FPGA bitstream for deployment. A front-end converts a Hardware Description Languate (HDL) design into a gate-level Register Transfer Level (RTL) netlist called as the logical synthesis step. This is subsequently passed through various back-ends of the EDA design flow to generate a bitstream which is then flashed onto hardware. The general aim is to always optimise the flow keeping a subset of Key Performance Indicators (KPIs) as their end target.  
However as effective as this flow is, a lot of optimisation pointers can be inferred from functional simulation to guide the back-end (place and route) tools. This is a common practice in compiler technologies for CPUs and GPUs which typically rely on designated entry points within the codebase to initiate translation from high-level programming languages to executable machine code. These entry points serve as pivotal anchors for the compiler's analysis and optimization processes, allowing for targeted optimizations tailored to specific runtime environments. The integration of runtime optimizations further enhances the efficiency and performance of compiled code, dynamically adjusting execution behavior based on runtime conditions and system characteristics. 
FPGA tools however are unable to leverage runtime optimisations used in CPUs and GPUs due to the following reasons:
\begin{itemize}
    \item \textbf{Hardware execution: } Runtime optimizations in traditional compilers for CPUs and GPUs focus on modifying software instructions or data layouts at runtime to improve performance. However, FPGAs execute hardware descriptions, which are configured before runtime and cannot be modified dynamically during execution like software instructions.
    \item \textbf{Configuration and constraints: }FPGAs are configured with a specific hardware design during configuration time, through a bitstream generated by a synthesis tool and remain fixed (unless there is a reconfiguration). Furthermore, FPGAs require careful resource allocation and optimization during synthesis and place-and-route stages to meet timing constraints and fit within the available hardware resources, unlike CPUs and GPUs with more flexible resource management.
\end{itemize}

There is also a notable distinction in the approach taken by FPGA tools toward the Intermediate Representation (IR) of HDL code. Any tool-flow pass is done on the IR with the inputs and outputs of an HDL module always being used to the maximum extent. Any dead-code elimination cannot be done on these ports making such a pass not effective on the code. These factors, combined with the fact that behavioural simulation is only limited to functional testing, lead to missed opportunities to perform runtime optimisation. We can visualise this in the code snippet  \ref{code:if_statement}. Assume the snippet to be run first to be run for \verb+31'hFOOBA clk+ cycles and secondly for \verb+31'hFOOB9 clk+ cycles. 
Also assume the input ports \verb+in+ receive values from the external interface (random conditions), \verb+reset+ is always set to high and \verb+clk+ keeps toggling edge every cycle. 
In the first scenario (of \verb+31'hFOOBA' clk+ cycles), entity \verb+pointless+ is useful as the output port vector (\verb+out+) is written to with a value (at line 16). In the second scenario (of \verb+31'hFOOB9' clk+ cycles), however, even though \verb+pointless+ would have no use as the scenario only runs for \verb+31'hFOOB9+ cycles, (1 cycle short from the line 15 condition being met).
If in the real-world use case, the second scenario is the most frequently evaluated condition and if we want our design to have the highest performance (timing), then the toolflow pass should be able to "speculate and optimise" such pathways in the final design to improve the latency of the datapath.    
\begin{verilog}[
    caption=\centering \small A simple conditional Verilog model which increments every clock cycle,
    label=code:if_statement,
  ]
// ports:
// input[31:0] in, reset, clk;
// output[31:0] out;
//
reg [31:0] count_c;    // assume = 0
reg [31:0] pointless;  // assume = 0 
always_ff @ (posedge clk) begin
    count_c <= count_c + 1;
    if (~reset) begin
        assign out = in;
    end
    else begin
        assign out = count_c;
        if (pointless == 31'hFOOBA) begin
            assign out = 31'hFOOBA;
        end
output[31:0]    end
end
always_ff @ (posedge clk) begin
    pointless <= pointless + 1;
end
\end{verilog}

In this paper, we present Simopt, a framework which uses the metadata generated from a behavioural simulation to speculatively optimise the HDL code to perform latency-driven placement and packing optimisation of user logic. 
Our specific contributions are as below:
\begin{itemize}
    \item Develop a framework for capturing important metadata from behaviour simulation, using an open-source simulator, Verilator.
    \item Integration of the metadata in an open-source placement tool, Yosys, to perform guided placement optimisation
    \item Benchmark circuits to quantify latency gains that can be achieved using the proposed framework.
\end{itemize}
We benchmarked the design using compact Verilog circuit models that are openly available through Verilator and EPFL benchmarks to validate the tool and quantify the performance gains.
Our results show that using the Simopt-pass in the design implementation flow results in up to 26.1\% reduction in timing performance (latency) of the design, and 5-10\% speed-up in the routing across multiple test designs.

\section{Background}
To the best of our knowledge, using behavioural simulation to accelerate a toolflow has not been widely explored in both CAD toolflow and compiler literature. The most analogous representation of this concept found in previous studies is from the FPGA design optimisation using roofline models \cite{siracusa2021comprehensive}. The authors used roofline models to highlight the efficient memory access patterns in particle methods \cite{del2018scalable}, wavefront algorithms \cite{haghi2023wfa}, and sparse arithmetic computation \cite{jain2020domain} achieving significant performance enhancements.
This however was very specialised to neural networks and used an LLVM IR to generate the best directives for their roofline models and does not adapt them into or use them as proxies to guide existing optimisation passes available in the EDA toolflow chain to generate a correspondingly optimised hardware. Similar works have been proposed in  \cite{zeng2022fpga} \cite{mohaidat2024survey}, where an inference algorithm generates an equivalent to an FPGA/ASIC design; however, these are domain-specific and rarely make any changes to the CAD toolchain. Whereas on the other side in the spectrum of academic literature, authors of \cite{kloibhofer2021run} devised a speculative compiler optimisation for data analysis which encodes multiple data structure optimisations (such as prioritise highly selective predicates, vectorisation). However this and other works \cite{sterling2017survey} focus mainly on compilers for CPUs and allow these concepts to be integrated into GPU compilers too. In this work we focus on FPGA/ASIC CAD tool flow which takes the metadata of a behavioural simulation as input, to generate a low-latency implementation.

\section{Simopt-Verilator}
\begin{figure}[t]
    \centering
    \includegraphics[scale=0.3]{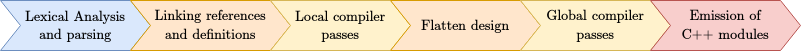}
    \caption{\centering \small  Traditional Verilator flow from a Verilog HDL to a C++ behavioural simulation. Verilator coins this compilation flow as \textit{verilate}. } 
    \label{fig:verilator_flow}
\end{figure}

The notion of using simulation metadata to guide low-level optimisations is driven by extensive functional simulations that are performed on complex designs. 
For the development of Simopt, we chose the popular open-source simulator, Verilator~\cite{verilator}.
Verilator generates a C++/SystemC behavioural model by compiling a Verilog (a HDL) design into a host-machine compatible executable. To run the simulation - a user needs to provide a C++ written wrapper (containing a standardised \verb+main()+) which is executed by the "simulation runtime". 
The main flow of the Verilator is shown in Fig. \ref{fig:verilator_flow}. 
\begin{figure}
    \centering
    \includegraphics[scale=0.4]{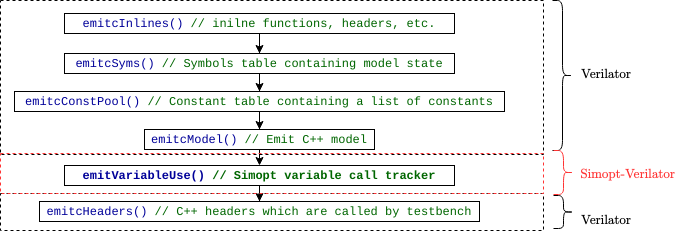}
    \caption{\centering \small Emission of the Verilator AST to C++ behavioural simulation steps} 
    \label{fig:verilator_emit}
\end{figure}

Verilator uses third-party tools (for lexical analysis and parsing)
to generate the Abstract-Syntax Tree (AST). This is then used by the subsequent step of the Verilator flow to link references (such as functions, and variables) to their definitions. After running a few local compiler passes, the AST is pseudo-flattened with each module scoped out. This makes it easier to run the final two steps on the optimised AST which is globally scoped compiler passes and emission of C++ modules. It is in the Verilator step of emission where the Simopt comes into play. This entire compilation (from Verilog to C++) is termed as the \textit{verilate} process in Verilator.   

Emission of the C++ modules is sequenced as shown in Figure \ref{fig:verilator_emit} illustrating Verilator's flow. Simopt-Verilator does an addition to the usual flow through the \verb+emitVariableUse()+ C++ class. The AST is processed by the 
Verilator internal emitter which first inlines any scoped-out modules, functions and headers. A symbol table is then generated containing various parameters of the model state (such as model precision, number of statements generated, model initialisation state and others). A constant table is also generated with a data structure of constants that have not been replaced by the constant propagation step (in the earlier Verilator compiler steps). The C++ model is subsequently generated using the emitted tables as a reference. 

To integrate Simopt into the Verilator simulator (through \verb+emitVariableUse()+), a set of counters (called \textit{simopt-trackers}) should be integrated which tracks each signal in the Verilog description (single-bit and vectored \verb+wire, reg, port+ types). 
Tracking each signal is also designed to be optional allowing the designer to choose specific nets/entities or modules/sub-modules for optimisation. This also allows designers to trade-off simulation execution time in Verilator by tracking only the signals/modules of interest.
%
The simopt-counters corresponding to these signals are incremented each time they are activated by the verilated runtime, except for initialisation statements. 
For example, referring to the code in Listing~\ref{code:if_statement}, simopt-counters for all signals within the if--else construct (lines 9, 12) are generated by the tool, if tracking is enabled for them.
In a specific case, if the \verb+reset+ is asserted as logic \verb+HIGH+ from the start of the simulation, then only the simopt-counters for signals within the \verb+else+ condition (lines 12 -- 17) are executed. 
This conditional registered logic creates additional overheads using the above tracking logic, both to keep track of each condition and also to understand the change in state of the registered signal.  
To tackle this problem, \textit{simopt-state} is introduced which checks for mismatch before and after an operation is applied to a single-bit / vectored register reference to keep track of the signal across multiple branches. 
Tracking using state flow instead of multiple counters reduces the overhead on the Verilator runtime for registered logic.
In the case of vectored register logic, the constraints imposed by the HDL description are accordingly unrolled by verilator's optimiser. 
Activations on vectored entities are treated as operations on packed datatypes and are expanded into activity at the bit-level for incrementing the simopt-counters. 
A pseudocode representation of this \textit{mask-and-increment} logic for tracking \textit{simopt-state} and selective incrementing of \textit{simopt-counters} is shown in algorithm~\ref{alg:mask-and-increment}.

\begin{algorithm*}[h]

    \caption{\small Simopt's mask-and-increment algorithm, wrapped in function \textbf{\textit{maskAndIncrement()}}: Used to track changes in a data type $\bm{var}$}
    \label{alg:mask-and-increment}
\begin{algorithmic}[1]
\footnotesize
\Function{maskAndIncrement}{$var$}
    \If{$\bm{var}.\text{isSingleBit()}$} \Comment{single bit data types}
        \If{$(\bm{var}.\text{value()} \oplus \bm{var}.\text{simoptState()}) \neq 0$}
            \State $\text{unpackedSingleIncrement}(var, 0, 0)$
        \EndIf
    \ElsIf{$\text{var.isPackable()}$} \Comment{Packed vectored entities - can be unrolled}
        \For{$\text{index} \in 0 : \bm{var}.\text{size()}$}
            \If{$(\bm{var}[index].\text{value()} \oplus \bm{var}[index].\text{simoptState()}) \neq 0$}
                \State $\text{unpackedSingleIncrement}(var, \text{size}, \text{index})$
            \EndIf
        \EndFor
    \EndIf
\EndFunction

\Function{unpackedSingleIncrement}{$var, \text{size}, \text{index}$}
    \If{$\text{size} == 0$}
        \State $\text{var\_entity} = \text{var}$
    \Else
        \State $\text{var\_entity} = \text{var}[index]$
    \EndIf
    \If{$!(\bm{var\_entity}.\text{isPackable()})$} \Comment{Unpacked entity, create a mask and increment bitflips only} 
        \State $\text{mask} = (\bm{var\_entity}.\text{value()} \oplus \bm{var\_entity}.\text{simoptState()})$
        \State $\text{mask\_index} = \_\_builtin\_ffsll(\text{mask})$
        \While{$(\text{mask\_index} \neq 0)$}
            \If{$\text{mask\_index} \neq 0$}
                \State $\bm{var\_entity}.\text{SimoptCounter}[\text{mask\_index}]++$
                \State $\bm{var\_entity}.\text{SimoptState} \oplus= (1 \ll (\text{mask\_index} - 1))$ 
            \EndIf
            \State $\text{mask\_index} = \_\_builtin\_ffsll(\text{mask})$
        \EndWhile
    \Else \Comment{Single-bit entity, simple increment will do}
        \State $\bm{var\_entity}.\text{simoptCounter}++$
        \State $\bm{var\_entity}.\text{simoptState} = \bm{var\_entity}.\text{value()}$ 
    \EndIf
\EndFunction
\end{algorithmic}

\end{algorithm*}

When the parser in Verilator encounters a signal definition, it statically generates the bit width and type definition of the internal representation of the signal (referred to as $\bm{var}$ in the algorithm). Each $var$ generated during the Verilator compilation flow follows the same steps in the \textit{mask-and-increment} algorithm (algorithm~\ref{alg:mask-and-increment}):
\begin{itemize}
    \item the value of $var$ is checked for dissimilarity with the simopt-state using the XOR ($\oplus$) operator
    \item the simopt-counter is then incremented if the values do not match
    \item if the simopt-counter is updated, update the simopt-state with the new value of $\bm{var}$
\end{itemize}
In algorithm~\ref{alg:mask-and-increment}, the function \texttt{*.simoptCounter} is a reference to the simopt-counter, and \texttt{*.simoptState()} is a reference to the simopt-state of data-type $var$. 
The call \texttt{\textunderscore\textunderscore builtin\textunderscore ffsll(mask)} is a GCC compiler intrinsic, which returns the position of the least significant bit that is set to high in the $mask$ plus 1. 
\textit{Packed} vector is a Verilog construct where multiple signals or variables are compactly organized into a contiguous bit sequence. \textit{Unpacked} vectors denote a collection of signals or variables with separate bit representations. The \textit{mask-and-increment} algorithm determines $var$ is packable through the $*.isPackable()$ boolean which is provided by the Verilator framework. For tracking unpacked entities, the "unpacked value" of $var$ is treated as a mask and used for comparing against the $var$'s simopt-state (shown in line 15). Lines 16 to 20 update the simopt-counters and simopt-states of each entity if there has been any change in the $var$ value. This greatly reduces the dependencies when using iterations in the design and handles each generated data type (from Verilator corresponding to an HDL signal) as a special case.  

Once the Simopt components are generated by the emission stage (second last step in Fig.~\ref{fig:verilator_emit}, the rest of the flow remains unchanged. Verilator creates public functions of all the steps, including those for functions that define simopt-counters and simopt-states initialisation. Verilator also creates a public header for the Simopt-enabled flow, which can be included by a testbench for instantiating the C++ model to perform functional validation. 
When performing functional validation, Simopt-enabled flow invokes the mask-and-increment algorithm for each internal variable in the verilated design (corresponding to the signals in the HDL) referenced by the runtime. 
Using the same example from listing \ref{code:if_statement}, Simopt-enabled Verilator flow will emit an executable as shown in listing~\ref{code:generate_if_statement}.

\begin{verilog}[
    caption=\centering \small Conditional Verilog model with emitted Simopt's mask-and-increments (function \textit{maskAndIncrement() } covered by algorithm \ref{alg:mask-and-increment} ),
    label=code:generate_if_statement
  ]
reg [31:0] count_c, pointless;
always_ff @ (posedge clk) begin
    count_c <= count_c + 1;
    maskAndIncrement(count_c);
    if (~reset) begin
        assign out = in;
        maskAndIncrement(in, out);
    end
    else begin
        assign out = count_c;
        if (pointless == 31'hFOOBA) begin
            assign out = 31'hFOOBA;
        end
        maskAndIncrement(count_c, out, pointless);
    end
end
maskAndIncrement(reset);
\end{verilog}

\begin{figure}[t]
    \centering
    \includegraphics[scale=0.29]{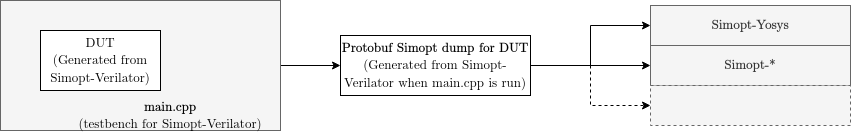}
    \caption{\centering \small The Simopt framework: Plug-and-use feature. Here \textit{Simopt-*} can be any other framework which has integrated the Simopt encoded dump.} 
    \label{fig:plug_and_use}
\end{figure}

Clock signals (\verb+clk+) are not tracked through Simopt and simopt-counters of such signals are assigned a maximum value (\verb+UINT128_MAX+) to assert their importance in the subsequent Simopt-driven placement optimisation stages. Synchronous \verb+reset+, on the other hand, is tracked by the Simopt logic since its activation has a bearing on the outputs of the system (lines 7 and 12 in listing \ref{code:generate_if_statement}). 
Once the functional simulation is completed, a dump of all simopt-counters with their corresponding (Verilog) variable names is generated before invoking the destructor function. To enable seamless integration of Simopt framework to any Verilator project, we use protocol buffers, specifically Protobuf due to their lower (de)compression times and memory footprint~\cite{serialisation_comp}, to handle serialisation of the dump. 
This enables the encoded dump to be used by other FPGA tools in the CAD flow as shown in figure \ref{fig:plug_and_use}. We discuss specific integration with Yosys in the next section (sec. ~\ref{sec:backends}) 

\section{Simopt-backends}\label{sec:backends}
For each design under test (DUT), we generate a Protobuf dump containing simopt-counters for each data type. Vectored entities (both packed and unpacked) are flattened with indices appended to the signal name. 
We use open-source Yosys tools to show the integration of simopt dump and how it can aid in improved mapping and placement of logic. 

Yosys provides a versatile environment for hardware description language (HDL) processing and synthesis tasks while allowing the functionality of the toolchain to be extended by user-defined optimistions~\cite{barzen2023narrowing}. 
Technology mapping in this flow is enabled by Berkeley's ABC tool~\cite{mishchenko2007abc}.
The technology mapping step optimises the synthesized netlist of the design and fine-tunes it for the target architecture, using advanced algorithms to perform logic minimisation and mapping of resulting logic to resources on the FPGA. 
These optimisations and mapping decisions have a significant impact on total resource consumption and the performance that can be achieved for a user design. 


We first explain the standard Yosys-ABC flow to show how simpot\_counters can be integrated to guide placement within this flow. 
First, complex hierarchical structures that are generated when converting Verilog to And-Inverter Gate (AIG) logic by the Yosys synthesis flow is flattened. 
Subsequently, technology mapping uses internal Verilog reference modules to substitute equivalent logic in the user design, followed by another flattening step before applying mapping optimisations. 
At this phase, the Protobuf files from the Simopt framework are deserialised to generate a map of netlist names with corresponding simopt counter values. 
To do this, a new backend script is generated that is invoked from within the Yosys flow to generate the simopt\_counter embedded flattened netlist file.
This netlist is used by the modified ABC flow where the initial steps (structural hashing, logic minimization, structural correction) are applied as-is to optimise the design and to perform technology-aware synthesis. 
Simopt-counters are integrated into the flow during the technology mapping stage (\verb+if+ command), which maps and optimises the design by identifying and factoring out common subexpressions across multiple levels of logic. 
At this stage, the tool determines the cost function for mapping the logic to Lookup tables (LUTs) using ``priority cuts'' where user-driven priority can be imposed. 
To integrate simopt-driven speculation, we optimise the priority cuts using the area cost function. 
The standard area cost function LUT mapping in ABC is formulated in equation \ref{eq:1}
\begin{equation}\label{eq:1}
    \small
    A = \sum_i^N\bigg(\frac{L_{inputs}}{L_{max\_inputs}}\times L_{outputs}\bigg)
\end{equation}
where $A$ is the LUT area, $L_{inputs}$ is the number of inputs to a logic design, $L_{max\_inputs}$ is the maximum number of inputs a LUT can have and $L_{outpts}$ is the number of outputs to a logic design. 
The function effectively aggregates individual area contributions ($i$ to $N$) to estimate the synthesized logics' area. 
The goal of this optimisation is to determine the mapping configuration(s) that minimises $A$.

To enable weighting using simpot\_counters, we introduce a logarithmic scaling factor for each design as shown in the equation~\ref{eq:2}. 
Fig~\ref{fig:log} shows the dampening effect we aim to achieve using the logarithmic scaling, allowing nets with lower count values to be assigned a lower weighting in the area minimisation logic, whereas nets with high count values (higher activations during simulation run) will be assigned a higher preference. 
The logarithmic scaling attaches a small bias to the area optimisation algorithm allowing logic elements driving and/or receiving the signals with high simopt\_counter activations to be preferentially packed and placed in this phase. 

\begin{equation}\label{eq:2}
    \small
    \begin{multlined}
    A_{simopt} = \sum_i^N\bigg[\frac{L_{inputs}}{L_{max\_inputs}} \times {} \\
    {} L_{outputs} \times \color{black}{\bigg(log\bigg(\frac{simopt\_score}{1+simopt\_score}\bigg)+1\bigg)}\color{black}{\bigg]}
    \end{multlined}
\end{equation}
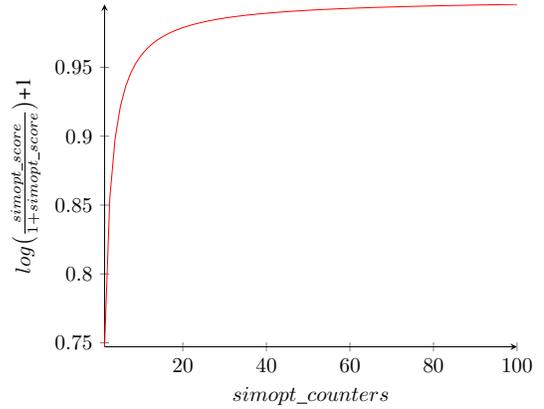
\begin{figure}
\centering
\begin{tikzpicture}[scale=0.8]
\begin{axis}[
    axis lines = left,
    xlabel = \(simopt\_counters\),
    ylabel = {\(log\big(\frac{simopt\_score}{1+simopt\_score}\big)\)+1},
]
\addplot [
    domain=0:100, 
    samples=80, 
    color=red,
]
{(log10(x/(x+1)))+1};

\end{axis}
\end{tikzpicture}
\caption{\centering \small  A graph of logarithmic scale factor vs the simopt\_counters. Higher the simopt\_counter, lower the variance in the scalar factor, allowing all high activations to be treated with importance}
\label{fig:log}
\end{figure}
Once the packing and placement are determined, a mapped netlist is generated in the subsequent phase and the Yosys tool is configured to generate a placed netlist in the Berkely Logic Interchange Format (BLIF) format. 
For our evaluation, we import this as a pre-placed IP package into AMD's Vivado tool with an AXI wrapper that integrates this design as an AXI slave IP on a Zynq platform for testing the performance. 
The implementation runs on Vivado are modified to treat the pre-placed IP as a black-box to ensure that the placement and routing generated through the simopt-enabled Yosys flow are preserved.
The generated design is subsequently deployed on a Pynq-Z2 block and its input-output latency is quantified using a FreeRTOS C driver on the ARM cores using standard register read/write operations. 


\section{Results} \label{sec:result}
For testing this framework, our target user designs have to be compatible with Verilator and are hence restricted to synthesisable Verilog 2001 designs that use constructs supported by Verilator. 
We used the internal Verilator benchmarks and open-source EPFL circuit benchmarks in this work to quantify the performance gains. 
The latency of the design was chosen as the parameter since the benchmark circuits are small designs and would not offer significant gains in resource consumption. 
We compare two versions of the design: one generated by the proposed simopt-enabled Yosys-ABC-Vivado flow and the second through a unmodified Yosys-ABC-Vivado flow.
All Verilator simulations were performed on a machine with AMD Ryzen-2700X CPU and 64\,GB of RAM. 
The FreeRTOS driver that interacts with the design sends input combinations to the deployed design and times the latency of the logic using the internal timer library that uses the timer blocks on the ARM core (running at 650\,MHz). 
An interrupt handler registered to the interrupt line of the wrapped AXI block is used to detect the completion of the operation by the block under test. 
Timestamps at the start and end are used to evaluate the elapsed time and hence the latency incurred by the block. 
The measurements are averaged over 1000 runs to ensure that any OS-level variability is captured and eliminated. 
 We summarise our results across the verilator\_ext\_tests test suite \cite{verilator-testsuite} in table \ref{table:result_yosys} and EPFL benchmarks (arithmetic section) \cite{amaru2015epfl} in table \ref{table:result_yosys2}.

\begin{table}[htbp]
\begin{center}
\begin{tabular}{@{}lccccc@{}}
    \toprule
	\textbf{Circuit} & \textbf{Simopt Time} & \textbf{Vanilla Time} & \textbf{\% savings} \\
        &  (ms) &  (ms) & (latency) \\
	\midrule
    t\_math\_wallace & 152.825 & 162.229 & \textcolor{darkgreen}{5.8}\\
    t\_synmul\_mul & 120.064 & 126.572 & \textcolor{darkgreen}{5.1}\\
    t\_math\_cond\_huge & 220.358 & 230.452 & \textcolor{darkgreen}{4.4}\\
    t\_wbuart32\_linetest & 149.93 & 157.98 & \textcolor{darkgreen}{5.1}\\
    \bottomrule
\end{tabular}
\caption{\centering \small Result of using Simopt-pass on in-built Verilator circuits}
\label{table:result_yosys}
\end{center}
\end{table}

\begin{table}[h]
\centering
\begin{tabular}{@{}lcccccc@{}}
	\toprule
	\textbf{Circuit} & \textbf{Simopt-Time} & \textbf{Vanilla-Time} & \textbf{\% savings}\\
                        &  (ms) &  (ms) & (latency) \\
	\midrule
    adder & 0.21 & 0.34 & \textcolor{darkgreen}{38.2} \\
    div & 4.76 & 5.08 & \textcolor{darkgreen}{6.3}\\
    hyp & 5.98 & 6.53 & \textcolor{darkgreen}{7.9}\\
    max & 0.31 & 0.43 & \textcolor{darkgreen}{27.9}\\
    sin & 1.89 & 2.28 & \textcolor{darkgreen}{17.1}\\
    multiplier & 1.36 & 1.42 & \textcolor{darkgreen}{4.2}\\
    sqrt & 2.87 & 3.12 & \textcolor{darkgreen}{8.0}\\
    square & 2.05 & 2.21 & \textcolor{darkgreen}{7.2}\\
	\bottomrule
    
\end{tabular}
\caption{\centering \small  Latency results of Simopt framework of the arithmetic circuits in EFPL benchmark}
\label{table:result_yosys2}
\end{table}

In tables~\ref{table:result_yosys} and~\ref{table:result_yosys2}, Simopt-Time corresponds to the latency measured for the design in which our simopt-driven optimisation was applied using our modified Simopt-Yosys-ABC and the resultant BLIF IP is used in the generated bitstream, with Vanilla-Time capturing the latency of the standard Yosys-ABC generated design. 
The benchmark design packaged within Verilator includes low-level primitives such as Wallace tree multiplier having inouts with bit-widths ranging from 16-bit to 128-bit connecting to sequential circuits. 
The results in table~\ref{table:result_yosys} show that simopt-driven optimisation can improve the circuit latency by at least 4.4\% across the larger benchmark circuits that are part of the Verilator tools. 

For the EPFL benchmark circuits shown in table~\ref{table:result_yosys2}, a fixed input value was used across the tests by varying the precision according to the input width of the circuit. 
The benchmark contains circuits with a minimum bit-width of 32 bits and a maximum of 128 bits. 
In the case of higher bit-widths, the inputs are padded to the required precision through sign extension or by zero-padding. 
Among the EPFL benchmarks, latency measurements for some cases such as \emph{bar} fell within the margin of measurement error for our setup (20\,\textmu s tick-rate of FreeRTOS) and are hence not shown in the table. 

From the results, it can be observed that Simopt-driven speculative optimisation can provide substantial improvements in the operating speed and latency of circuits, particularly in the case of combinational logic. Moreover, there was no observed change in FPGA utilisation area running the benchmarks through the Simopt framework.
It should, however, be noted that additional logging required for Simopt metadata generation does incur longer simulation runtimes on Verilator. 
With large combinational designs, we observed that logging all signals can cause up to 5$\times$ increase in simulation time in Verilator, and further user-driven optimisations could be explored in this case. 
The implementation and runtime performance in the Simopt-Yosys-ABC flow was insignificant, with a worst-case of 5\% increased runtime that was observed in our tests. 

\section{Conclusion}
In this work, we demonstrated the case for using metadata generated during the circuit design and validation phase (through behavioural simulations) to guide the optimisations during the placement and routing phases of the CAD flow. 
The Simopt framework introduced in this work is a plug-and-play model that can be adapted to any simulation framework to generate the \textit{simopt-dump}, to be fed into the later place \& route tools to generate lower area or lower latency models. 
In the use case discussed in this paper, we demonstrated the flow using an open-source Verilator tool for simulation and the open-source Yosys-ABC tool for synthesis, place \& route, adapting the CAD flow to utilise simulation metadata for altering the priority of cuts during the placement flow.
The integrated flow was benchmarked using openly available circuit descriptions with the results showing that substantial improvements to the circuit's latency can be achieved using this flow. 
In the future, we propose to investigate methods to further increase area and/or latency savings by applying optimisations across multiple hierarchies and to the routing flow, while also investigating methods to reduce overheads incurred during the metadata generation phase. 

\bibliographystyle{IEEEtran}  
\bibliography{sample-base}  

\end{document}